# PREPROCESSING IN ATTRACTOR NEURAL NETWORKS

C.G. CARVALHAES, A.T. COSTA JR. and T.J.P. PENNA

*Instituto de Física, Universidade Federal Fluminense*

*C.P. 100296, 24001-970, Niterói, RJ, Brazil*

Preprocessing the input patterns seems the simplest approach to invariant pattern recognition by neural networks. The Fourier transform has been proposed as an appropriate and elegant preprocessor. Nevertheless, we show in this work that the performance of this kind of preprocessor is strongly affected by the number of stored informations. This is because the phase of the Fourier transform plays a more important role than the amplitude in the recognition process.

## 1. Introduction

Neural networks have been extensively studied during the last years and most of the work done along this line has been devoted to the associative memory features emerging from these systems[1]. Since associative memory devices permit the recall of information from a stimulus which, by its turn, can contain only a fraction of the total information, they are frequently used with pattern recognition purposes. Some remarkable examples of neural networks used in this task are the NEOCOGNITRON [2], proposed by Fukushima to recognize handwritten numerals and the NET-TALK [3] which reads a text written in English and is able to pronounce it correctly with an accuracy of around 80% when new words are presented. Although these examples have fast response they require an exhaustive learning, characteristic of multilayer networks executing complex tasks.

A special class of neural network topology is that of fully connected networks, in which the most studied model is the Hopfield model[4]. For retrieval of information in this sort of neural networks, one of the crucial parameters is the Hamming distance between the stimulus and the stored information. If this distance is small, then the network relaxes to the correct information. Due to this fact they are sometimes called "attractor neural networks". One advantage of this class of neural network concerns the learning step that is easier and faster in this topology than in feed-forward networks. Besides these considerations, the recognition of optical patterns subject to transformations like translations, rotations and scaling, for example, is a very difficult task for the present neural network models (for both feed-forward and fully connected topologies) due to the high degree of nonlinearity introduced in these transformations. The simplest way of dealing with invariant patterns is the preprocessing. Strong evidences for preprocessing in visual systems of superior animals have been suggested [5]. Among some proposals of preprocessors[6-9], maybe the most elegant and promising is the one using the Fourier transformation as studied by Fuchs and Haken[7] in a content-addressable memory consisting of nonlinear units[10]. Since Fourier transform is easily implemented in massively parallel computers, this preprocessor seemed to be adequate for this task. In this work we present results from extensive numerical simulation in large fully-connected neural networks which show that the use of preprocessor for practical situations (for example, a large number of stored informations) can be very inefficient in opposition to current ideas.

This paper is organized as follows: in Section 2 we review a recent powerful model for attractor neural networks, in Section 3 we describe the Fourier preprocessor as prescribed by Fuchs and Haken[7] and, in Section 4, our Boolean version of this preprocessor. The numerical results are presented in Section 5 and finally some remarks and conclusions.

## 2. The RS model

A neural network is an assembly of simple computing elements (neurons) which, in the simplest models, may be represented by Ising spins, i.e., the state of a given neuron $i$, at time $t$, is given by $S_i(t) = \pm 1$. The strength of the connections $J_{ij}$ (synaptic efficacies) between the neurons $S_i$ and $S_j$ will determine which informations will be stored in the network. This system can be used as a device to recognize scanned black and white pictures (hereafter called *"patterns"*) where for each black pixel we associate a variable $\xi_i = +1$ and for the white pixels $\xi_i = -1$.

Recently a new model for neural networks has been proposed: the $RS$ model[11,12]. The new energy function for a $N$-neuron network, with $P$ stored patterns, is given by:

$$\mathcal{H} = N \prod_{\mu=1}^{P} \left[ \frac{1}{2N} \sum_{i=1}^{N} (\xi_i^\mu - S_i)^2 \right] \tag{1}$$

or equivalently,

$$\mathcal{H} = N \prod_{\mu=1}^{P} (1 - m_\mu), \tag{2}$$

where $m_\mu$ is the overlap between the current state of the network and a given pattern $\xi^\mu$, defined as

$$m_\mu = \frac{1}{N} \sum_i \xi_i^\mu S_i . \tag{3}$$

It is clear that the global minima of this energy function ($\mathcal{H} = 0$) correspond to $m_\mu = 1$ one particular pattern $\mu$. Hence all the patterns will be learned by the network independently of the number and of the correlation between them, overcoming some restrictions of the original Hopfield model[4]. In this version the anti-memories ($\xi^\nu \equiv -\xi^\mu$) are not minima of eq.(1). Numerical simulations have shown that when the anti-memories are explicitly stored [13,14] (**patterns and anti-patterns stored version**), the performance of this model is greatly improved, mainly in the storage of correlated patterns. The explicit learning of the anti-memories can be implemented by rewriting (2) as

$$\mathcal{H} = N \prod_{\mu=1}^{P} (1 - m_\mu^2). \qquad (4)$$

If we expand the energy function (4) the multineuron feature becomes evident and we find terms up to the $P^{th}$ order in the synapses

$$\mathcal{H} = N \left( 1 - \sum_{\mu_1} m_{\mu_1}^2 + \sum_{\mu_1 < \mu_2} m_{\mu_1}^2 m_{\mu_2}^2 - \cdots + (-1)^P \sum_{\mu_1 < \cdots < \mu_P} m_{\mu_1}^2 \cdots m_{\mu_P}^2 \right). \qquad (5)$$

The first non-trivial term is the Hopfield energy function. The multineuron interaction terms behave as corrections to the Hopfield model. There are strong evidences for multineuron interactions in biological systems[15]. Moreover, high-order synapses have been proposed as one of the ways to introduce nonlinearity and to solve the invariant pattern recognition in feed-forward neural networks[16]. It has been shown[15-19] that high-order synapses also improve the capacity of neural networks concerning the number of stored patterns. Nevertheless, there is nothing to support the existence of an extensive order in the synapses. The effects of truncation of equation (5) up to the fourth order have been studied elsewhere[18,19], and it has been shown that the truncated model (TRS) can store an extensive number of patterns ($\mathcal{O}(N^3)$) for an optimal weight of the fourth order term. Due mainly to the simplicity of the RS model and its more powerful performance for finite sizes we choose to work with it instead of the TRS model. Let us stress here that the TRS is easier to be implemented in hardware and more realistic from a biological point of view.

### 3. Preprocessing the Input Patterns

As we pointed out in section 2, attractor neural networks are not capable of invariant perception although real brains can recognize objects independently of position, scale and angle of vision with relative facility. In this section we review a proposal of a preprocessor, studied by Fuchs and Haken[7] in a non-neural system[10], which uses Fourier transformations to overcome these difficulties. Other non-neural approaches and comparisons with multilayer networks for recognition of boundaries can be found in the literature[20].

Let us consider a two-dimensional function $s(x,y)$. This function can be Fourier transformed, i.e. represented in the $k$-space, as

$$\tilde{s}(k_x, k_y) = \int \mathrm{d}x \int \mathrm{d}y \; \exp(\mathrm{i}k_x x + \mathrm{i}k_y y) \, s(x,y). \qquad (6)$$

4 *C.G. Carvalhaes, A.T. Costa Jr. & T.J.P. Penna*The quantity $|\tilde{s}(k_x, k_y)|$ will be invariant against translations in the plane. In polar coordinates we have

$$(k_x, k_y) = (e^q \cos\phi, e^q \sin\phi). \tag{7}$$

Rotations are equivalent to a shift in the polar angle, while a scale transformation correspond to shift the $q$–coordinate. Applying a second Fourier transformation, we have

$$\hat{s}(u,v) = \int dq \int d\phi \exp(iuq + iv\phi) |\tilde{s}(e^q \cos\phi, e^q \sin\phi)|, \tag{8}$$

and $|\hat{s}(u,v)|$ is now invariant against all three transformations: translation, rotation and scaling.

## 4. Binary Version of the Preprocessor

As pointed before, Fuchs and Haken studied Fourier preprocessing in a pattern recognizing device with synergetic units[10] that are continuous variables. Since we intend to study this preprocessing in Ising-like neural networks, the development of a discrete version of it is needed. The advantages of treating Boolean variables are evident, they save memory and computer time. The two-dimensional discrete Fourier transform of a $M \times N$ picture can be written as

$$\tilde{s}(k_x, k_y) = \frac{1}{MN} \sum_{x=0}^{M-1} \sum_{y=0}^{N-1} \exp\left[-2\pi i \left(\frac{k_x x}{M} + \frac{k_y y}{N}\right)\right] s(x,y). \tag{9}$$

Many methods of calculating discrete Fast Fourier Transform (FFT) have been proposed and developed in the last years[21]. We used in this work the successive doubling strategy.

By checking eq.(9) we note that $|\tilde{s}(k_x, k_y)|$ is not a discrete variable. In order to work with boolean variables, which are discrete, we proceed as follows: we scale $|\tilde{s}(k_x, k_y)|$ conveniently and discretize it, so that it can only assume integer values between 0 and $n-1$. To each point $(k_x, k_y)$ in $k$-space it will correspond a square with $n$ $k$-space neurons. The discrete value of $|\tilde{s}(k_x, k_y)|$ will be the number of active $k$-space neurons in the square. Every $n$-$k$-space neuron square in the preprocessed pattern with a given activity will have the same pattern of active neurons. For example, we choose that the only active $k$-space neuron in a square with activity one will be the third. Then to every point in $k$-space for which $|\tilde{s}(k_x, k_y)| = 1$ it will correspond a square with only the third neuron active. This procedure is different from the one used by Chen *et al.*[22] to discretize diffraction patterns. There the pattern of the $k$-space squares were chosen randomly at each point.

Since we have just configured our preprocessor we must test how load capacity and sizes of basins of attraction are affected by the preprocessing and eventually by the double transformation needed for rotation and scale invariance. In the work of Fuchs and Haken[7] only the retrieval of one in five pattern was shown. This is not sufficient to give conclusive results about the points mentioned above. Here we intend to present a quantitative study of this problem by performing numerical simulations. The results are presented in the next section.

## 5. Results

Firstly, we present the results considering only translations in the original patterns. Concerning the computational aspects of this work, we use the multineuron coding[23,24] and discrete FFT successive doubling algorithm[21,22]. The multispin coding is necessary since the preprocessed networks are large. The version of the multineuron coding for the RS model is presented in details elsewhere[14].

The size of the basins of attraction is one of the most important quantities in determining the performance of a neural network model. It can be obtained from numerical simulations through the retrieval curves. The retrieval curves, by their turn, are obtained measuring the fraction of correctly retrieved patterns. It is presented as input to the neural network a pattern which has a given initial overlap $m_i$ (eq.(3)) with a given stored pattern. If the equilibrium state of the network is exactly that stored pattern then we say that the retrieval was successfull. The fraction of correct retrievals $f(m_i)$ versus the initial overlap $m_i$, for a given number of stored patterns, is the retrieval curve.

At this point, it is worth explaining in details how the retrieval occurs with preprocessing. First, the stored patterns are the preprocessed ones as described in the last section. Hence, the dynamics will be applied in the $nN$-neuron network. For practical purposes, the crucial information is whether the original pattern was (or not) recognized by the neural net. Therefore, when we refer to initial overlap in this work we mean the overlap between the *original* patterns. This is important because we might apply the anti-transformation to recover the original pattern. However, for testing recognition, we can compare the output (equilibrium state) of the neural net with the stored patterns or use some neurons as index for each pattern[9] using, for example, the binary representation of the number of the pattern (no comparison is needed in this treatment). It is also important to remember that, since the information about the phase is lost in taking the modulus of the Fourier transformation, the shift in the original patterns is irrelevant for this kind of preprocessor.

In figure 1, we present the retrieval curves for some values of the load parameter $\alpha = P/N$. These curves show the frequency of correct retrievals as a function of the initial overlap for an original $N = 256$ network transformed in a $N' = 2048$ network ($n = 8$). The values of $\alpha$ are taken considering the original size of the network, by convention. The important information is how the sizes of the basins of attraction are affected by the load parameter. This performance is better than the original model concerning *shifted* patterns. For any displacement, the retrieval is successfull, if no noise is included. However the performance is considerably worse than in the RS model for *non-shifted* uncorrelated patterns[13] (full symbols in fig.1). By the way, the main difference between the biological networks and Fourier preprocessor-based ones is that the larger the shift the more difficult is the recognition for the primer whereas in the latter the performance is exactly the same independently of the shift.

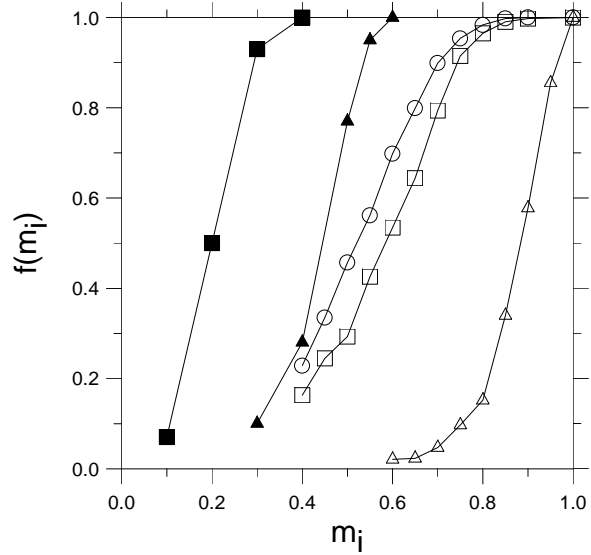

Fig. 1. Retrieval curves for different values of load parameter $\alpha$ for a $N = 256$ neural network: 0.05 (○), 0.1 (□) and 0.5 (△). For the sake of comparison we present the results for the RS model without preprocessing and *uncorrelated non-shifted* patterns, represented by the full symbols. The solid lines are guide to eyes.

The reason the performance is strongly affected by the Fourier transform is that the phase plays a more important role in caractherization of patterns than the amplitude. It is not a new conclusion[25] but here we could quantify this effect. The Fourier reconstruction belongs to a class of problems called "Inverse Problems" and small differences in the Fourier transforms can lead to very different functions when their respectives anti-transforms are taken. On the other hand, associative memory devices (like neural networks) can recover the complete information from a partial one. The question is whether attractor neural networks could be effective in retrieving this information from Fourier transformations. Unfortunately, this is not the case, according to fig.1. It is a fact that if the input is one of the patterns, using the RS model, the recognition is always successfull. However, even a small noise is enough to decrease this performance, since most of the information about the picture is lost in discarding the phase. In fact, the amplitude is more important to distinguish between *periodic* and *non-periodic patterns*.

In figure 2, we can see how the overlap between two patterns is affected by a Fourier transform. For a given initial overlap between random patterns $m_i$, we measured the mean final overlap $m_f$. The final overlap decreases exponentially as the initial overlap also decreases. For example, if the initial overlap between a pattern and the stimulus is $m_i = 0.8$, after the transformation the overlap will change to $m'_i = 0.4$. However, even taking into account this discrepancy, the original model will always recognize an information, at $\alpha = 0.4$, for $m_i = 0.4$ whereas if we include preprocessing it will not occur, as we can see in fig.1. This effect is due to the correlations introduced in the patterns. It is a well-known fact that correlations among the stored patterns decrease drastically the sizes of basins of attraction. We have also tested the double preprocessing for scaling and rotations. As expected the performance is still worse (represented in fig.2 by the dashed line). In this case the recognition only is successfull, in practical terms, if no noise is added. The curves present non-vanishing values for lower initial overlaps due to the zero-$th$ order term which contains information only about the activity. The region close to the central maximum gives origin to a macroscopic correlation between the patterns.

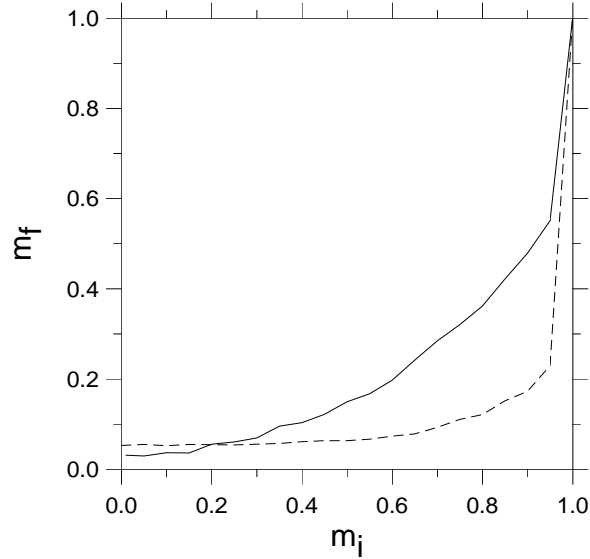

Fig. 2. Overlap between two patterns before ($m_i$) and after ($m_f$) the Fourier preprocessing (solid line). The overlap after two successive transformations is represented by the dashed line.

Another effect that could worse the performance of this process is the degree of discretization of the Fourier transform. In fig.3 we can see that it is not the case. The retrieval curves, for different levels of discretization $n$, collapse for $n =$ 4,9 and 16. We have not tested the preprocessor performance in the Haken's associative memory model, since it consists of units that can assume real (not integer) values. This model may present a better performance than the attractor neural networks presented here, but it includes a lack of simplicity that is not interesting for practical implementations. Moreover, our results corroborate the idea that the loss of the information carried by the phase is the main responsible for the bad performance of the preprocessing, by increasing the distance between the stimulus and the patterns and introducing correlations between the stored patterns.

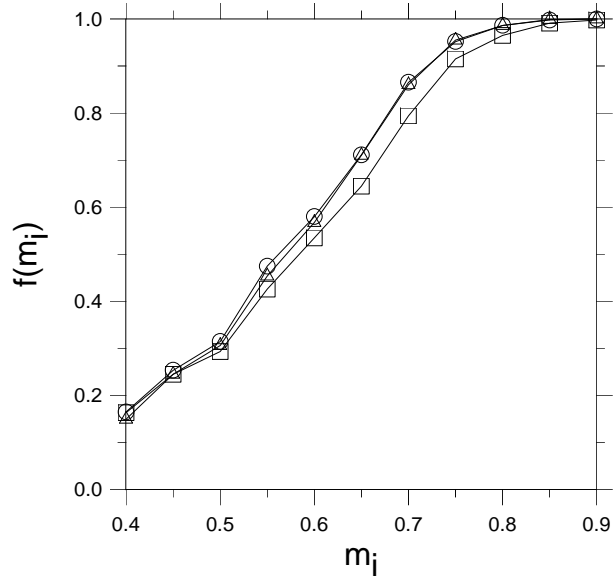

Fig. 3. Retrieval curves for different degrees of discretization $n=$ 4 (○), 9(□) and 16 (△).

## 6. Conclusions

Through numerical simulations we can show that Fourier transform preprocessor makes possible the recognition of shifted patterns even when adding some noise, however the performance is bad if the number of stored informations is large. This is because the phase of the Fourier Transform, that is lost in our preprocessing, contains much of the essential information about an image, and the intensity (the remanescent information) is very similar for all the patterns, mainly for the non periodic ones. We could quantify, using the overlap definition, how important the phase is. If we extend the preprocessor to recognize rotated or scaled patterns the situation will be more critical since the Fourier transform must be applied twice and consequently the stimulus will be even more distant from the stored patterns. In face of the results presented here, we suggest that even other models of attractor neural networks that present better performance with respect to storage of correlated patterns[26] will not strongly improve the results presented here. In particular, for the RS model, the implementation of dream sleep[27] could enhance the performance, since the effects of the central maximum in the transformation would be discarded. For the Hopfield model, the doubling strategy[28] could be effective if only the differences between the patterns are stored. Another possible (and probably more effective) solution could be the use of other invariant transformations than the Fourier Transform.

### Acknowledgements

We are grateful to S.M. Moss de Oliveira, P.M.C de Oliveira and J.J. Arenzon for useful discussions. Work partially supported by Brazilian agencies CNPq, CAPES and FINEP.